\documentclass[ reprint,superscriptaddress, amsmath,amssymb,aps,nofootinbib,bibnotes,floatfix]{revtex4-1}
\usepackage{graphicx}
\pdfoutput=1
\usepackage{epstopdf}
\usepackage{color}
\usepackage{dcolumn}
\usepackage{bm}
\usepackage{url}
\usepackage{upgreek} 
\usepackage{siunitx}
\usepackage{booktabs}
\usepackage[normalem]{ulem} 
\usepackage{placeins}
\usepackage{footnote}
\usepackage{hyperref}
\usepackage[utf8]{inputenc}

\newcommand{\GeVmass}{\ensuremath{\mathrm{GeV}\,c^{-2}}}

\newif\ifsection
\sectionfalse

\begin{document}

\title{Limits on spin-dependent WIMP-nucleon cross section obtained from the complete LUX exposure}

\author{D.S.~Akerib} \affiliation{Case Western Reserve University, Department of Physics, 10900 Euclid Ave, Cleveland, OH 44106, USA} \affiliation{SLAC National Accelerator Laboratory, 2575 Sand Hill Road, Menlo Park, CA 94205, USA} \affiliation{Kavli Institute for Particle Astrophysics and Cosmology, Stanford University, 452 Lomita Mall, Stanford, CA 94309, USA}
\author{S.~Alsum} \affiliation{University of Wisconsin-Madison, Department of Physics, 1150 University Ave., Madison, WI 53706, USA}  
\author{H.M.~Ara\'{u}jo} \affiliation{Imperial College London, High Energy Physics, Blackett Laboratory, London SW7 2BZ, United Kingdom}  
\author{X.~Bai} \affiliation{South Dakota School of Mines and Technology, 501 East St Joseph St., Rapid City, SD 57701, USA}  
\author{A.J.~Bailey} \affiliation{Imperial College London, High Energy Physics, Blackett Laboratory, London SW7 2BZ, United Kingdom}  
\author{J.~Balajthy} \affiliation{University of Maryland, Department of Physics, College Park, MD 20742, USA}  
\author{P.~Beltrame} \affiliation{SUPA, School of Physics and Astronomy, University of Edinburgh, Edinburgh EH9 3FD, United Kingdom}  
\author{E.P.~Bernard} \affiliation{University of California Berkeley, Department of Physics, Berkeley, CA 94720, USA} \affiliation{Yale University, Department of Physics, 217 Prospect St., New Haven, CT 06511, USA} 
\author{A.~Bernstein} \affiliation{Lawrence Livermore National Laboratory, 7000 East Ave., Livermore, CA 94551, USA}  
\author{T.P.~Biesiadzinski} \affiliation{Case Western Reserve University, Department of Physics, 10900 Euclid Ave, Cleveland, OH 44106, USA} \affiliation{SLAC National Accelerator Laboratory, 2575 Sand Hill Road, Menlo Park, CA 94205, USA} \affiliation{Kavli Institute for Particle Astrophysics and Cosmology, Stanford University, 452 Lomita Mall, Stanford, CA 94309, USA}
\author{E.M.~Boulton} \affiliation{University of California Berkeley, Department of Physics, Berkeley, CA 94720, USA} \affiliation{Lawrence Berkeley National Laboratory, 1 Cyclotron Rd., Berkeley, CA 94720, USA} \affiliation{Yale University, Department of Physics, 217 Prospect St., New Haven, CT 06511, USA} 
\author{P.~Br\'as} \affiliation{LIP-Coimbra, Department of Physics, University of Coimbra, Rua Larga, 3004-516 Coimbra, Portugal}  
\author{D.~Byram} \affiliation{University of South Dakota, Department of Physics, 414E Clark St., Vermillion, SD 57069, USA} \affiliation{South Dakota Science and Technology Authority, Sanford Underground Research Facility, Lead, SD 57754, USA} 
\author{S.B.~Cahn} \affiliation{Yale University, Department of Physics, 217 Prospect St., New Haven, CT 06511, USA}  
\author{M.C.~Carmona-Benitez} \affiliation{Pennsylvania State University, Department of Physics, 104 Davey Lab, University Park, PA  16802-6300, USA} \affiliation{University of California Santa Barbara, Department of Physics, Santa Barbara, CA 93106, USA} 
\author{C.~Chan} \affiliation{Brown University, Department of Physics, 182 Hope St., Providence, RI 02912, USA}  
\author{A.A.~Chiller} \affiliation{University of South Dakota, Department of Physics, 414E Clark St., Vermillion, SD 57069, USA}  
\author{C.~Chiller} \affiliation{University of South Dakota, Department of Physics, 414E Clark St., Vermillion, SD 57069, USA}  
\author{A.~Currie} \affiliation{Imperial College London, High Energy Physics, Blackett Laboratory, London SW7 2BZ, United Kingdom}  
\author{J.E.~Cutter} \affiliation{University of California Davis, Department of Physics, One Shields Ave., Davis, CA 95616, USA}  
\author{T.J.R.~Davison} \affiliation{SUPA, School of Physics and Astronomy, University of Edinburgh, Edinburgh EH9 3FD, United Kingdom}  
\author{A.~Dobi} \affiliation{Lawrence Berkeley National Laboratory, 1 Cyclotron Rd., Berkeley, CA 94720, USA}  
\author{J.E.Y.~Dobson} \affiliation{Department of Physics and Astronomy, University College London, Gower Street, London WC1E 6BT, United Kingdom}  
\author{E.~Druszkiewicz} \affiliation{University of Rochester, Department of Physics and Astronomy, Rochester, NY 14627, USA}  
\author{B.N.~Edwards} \affiliation{Yale University, Department of Physics, 217 Prospect St., New Haven, CT 06511, USA}  
\author{C.H.~Faham} \affiliation{Lawrence Berkeley National Laboratory, 1 Cyclotron Rd., Berkeley, CA 94720, USA}  
\author{S.R.~Fallon} \affiliation{University at Albany, State University of New York, Department of Physics, 1400 Washington Ave., Albany, NY 12222, USA}  
\author{S.~Fiorucci} \affiliation{Lawrence Berkeley National Laboratory, 1 Cyclotron Rd., Berkeley, CA 94720, USA} \affiliation{Brown University, Department of Physics, 182 Hope St., Providence, RI 02912, USA} 
\author{R.J.~Gaitskell} \affiliation{Brown University, Department of Physics, 182 Hope St., Providence, RI 02912, USA}  
\author{V.M.~Gehman} \affiliation{Lawrence Berkeley National Laboratory, 1 Cyclotron Rd., Berkeley, CA 94720, USA}  
\author{C.~Ghag} \affiliation{Department of Physics and Astronomy, University College London, Gower Street, London WC1E 6BT, United Kingdom}  
\author{M.G.D.~Gilchriese} \affiliation{Lawrence Berkeley National Laboratory, 1 Cyclotron Rd., Berkeley, CA 94720, USA}  
\author{C.R.~Hall} \affiliation{University of Maryland, Department of Physics, College Park, MD 20742, USA}  
\author{M.~Hanhardt} \affiliation{South Dakota School of Mines and Technology, 501 East St Joseph St., Rapid City, SD 57701, USA} \affiliation{South Dakota Science and Technology Authority, Sanford Underground Research Facility, Lead, SD 57754, USA} 
\author{S.J.~Haselschwardt} \affiliation{University of California Santa Barbara, Department of Physics, Santa Barbara, CA 93106, USA}  
\author{S.A.~Hertel} \affiliation{University of Massachusetts, Department of Physics, Amherst, MA 01003-9337 USA} \affiliation{Lawrence Berkeley National Laboratory, 1 Cyclotron Rd., Berkeley, CA 94720, USA} \affiliation{Yale University, Department of Physics, 217 Prospect St., New Haven, CT 06511, USA}
\author{D.P.~Hogan} \affiliation{University of California Berkeley, Department of Physics, Berkeley, CA 94720, USA}  
\author{M.~Horn} \affiliation{South Dakota Science and Technology Authority, Sanford Underground Research Facility, Lead, SD 57754, USA} \affiliation{University of California Berkeley, Department of Physics, Berkeley, CA 94720, USA} \affiliation{Yale University, Department of Physics, 217 Prospect St., New Haven, CT 06511, USA}
\author{D.Q.~Huang} \affiliation{Brown University, Department of Physics, 182 Hope St., Providence, RI 02912, USA}  
\author{C.M.~Ignarra} \affiliation{SLAC National Accelerator Laboratory, 2575 Sand Hill Road, Menlo Park, CA 94205, USA} \affiliation{Kavli Institute for Particle Astrophysics and Cosmology, Stanford University, 452 Lomita Mall, Stanford, CA 94309, USA} 
\author{R.G.~Jacobsen} \affiliation{University of California Berkeley, Department of Physics, Berkeley, CA 94720, USA}  
\author{W.~Ji} \affiliation{Case Western Reserve University, Department of Physics, 10900 Euclid Ave, Cleveland, OH 44106, USA} \affiliation{SLAC National Accelerator Laboratory, 2575 Sand Hill Road, Menlo Park, CA 94205, USA} \affiliation{Kavli Institute for Particle Astrophysics and Cosmology, Stanford University, 452 Lomita Mall, Stanford, CA 94309, USA}
\author{K.~Kamdin} \affiliation{University of California Berkeley, Department of Physics, Berkeley, CA 94720, USA}  
\author{K.~Kazkaz} \affiliation{Lawrence Livermore National Laboratory, 7000 East Ave., Livermore, CA 94551, USA}  
\author{D.~Khaitan} \affiliation{University of Rochester, Department of Physics and Astronomy, Rochester, NY 14627, USA}  
\author{R.~Knoche} \affiliation{University of Maryland, Department of Physics, College Park, MD 20742, USA}  
\author{N.A.~Larsen} \affiliation{Yale University, Department of Physics, 217 Prospect St., New Haven, CT 06511, USA}  
\author{C.~Lee} \affiliation{Case Western Reserve University, Department of Physics, 10900 Euclid Ave, Cleveland, OH 44106, USA} \affiliation{SLAC National Accelerator Laboratory, 2575 Sand Hill Road, Menlo Park, CA 94205, USA} \affiliation{Kavli Institute for Particle Astrophysics and Cosmology, Stanford University, 452 Lomita Mall, Stanford, CA 94309, USA}
\author{B.G.~Lenardo} \affiliation{University of California Davis, Department of Physics, One Shields Ave., Davis, CA 95616, USA} \affiliation{Lawrence Livermore National Laboratory, 7000 East Ave., Livermore, CA 94551, USA} 
\author{K.T.~Lesko} \affiliation{Lawrence Berkeley National Laboratory, 1 Cyclotron Rd., Berkeley, CA 94720, USA}  
\author{A.~Lindote} \affiliation{LIP-Coimbra, Department of Physics, University of Coimbra, Rua Larga, 3004-516 Coimbra, Portugal}  
\author{M.I.~Lopes} \affiliation{LIP-Coimbra, Department of Physics, University of Coimbra, Rua Larga, 3004-516 Coimbra, Portugal}  
\author{A.~Manalaysay} \affiliation{University of California Davis, Department of Physics, One Shields Ave., Davis, CA 95616, USA}  
\author{R.L.~Mannino} \affiliation{Texas A \& M University, Department of Physics, College Station, TX 77843, USA}  
\author{M.F.~Marzioni} \affiliation{SUPA, School of Physics and Astronomy, University of Edinburgh, Edinburgh EH9 3FD, United Kingdom}  
\author{D.N.~McKinsey} \affiliation{University of California Berkeley, Department of Physics, Berkeley, CA 94720, USA} \affiliation{Lawrence Berkeley National Laboratory, 1 Cyclotron Rd., Berkeley, CA 94720, USA} \affiliation{Yale University, Department of Physics, 217 Prospect St., New Haven, CT 06511, USA}
\author{D.-M.~Mei} \affiliation{University of South Dakota, Department of Physics, 414E Clark St., Vermillion, SD 57069, USA}  
\author{J.~Mock} \affiliation{University at Albany, State University of New York, Department of Physics, 1400 Washington Ave., Albany, NY 12222, USA}  
\author{M.~Moongweluwan} \affiliation{University of Rochester, Department of Physics and Astronomy, Rochester, NY 14627, USA}  
\author{J.A.~Morad} \affiliation{University of California Davis, Department of Physics, One Shields Ave., Davis, CA 95616, USA}  
\author{A.St.J.~Murphy} \affiliation{SUPA, School of Physics and Astronomy, University of Edinburgh, Edinburgh EH9 3FD, United Kingdom}  
\author{C.~Nehrkorn} \email{cnehrkorn@ucsb.edu} \affiliation{University of California Santa Barbara, Department of Physics, Santa Barbara, CA 93106, USA}  
\author{H.N.~Nelson} \affiliation{University of California Santa Barbara, Department of Physics, Santa Barbara, CA 93106, USA}  
\author{F.~Neves} \affiliation{LIP-Coimbra, Department of Physics, University of Coimbra, Rua Larga, 3004-516 Coimbra, Portugal}  
\author{K.~O'Sullivan} \affiliation{University of California Berkeley, Department of Physics, Berkeley, CA 94720, USA} \affiliation{Lawrence Berkeley National Laboratory, 1 Cyclotron Rd., Berkeley, CA 94720, USA} \affiliation{Yale University, Department of Physics, 217 Prospect St., New Haven, CT 06511, USA}
\author{K.C.~Oliver-Mallory} \affiliation{University of California Berkeley, Department of Physics, Berkeley, CA 94720, USA}  
\author{K.J.~Palladino} \affiliation{University of Wisconsin-Madison, Department of Physics, 1150 University Ave., Madison, WI 53706, USA} \affiliation{SLAC National Accelerator Laboratory, 2575 Sand Hill Road, Menlo Park, CA 94205, USA} \affiliation{Kavli Institute for Particle Astrophysics and Cosmology, Stanford University, 452 Lomita Mall, Stanford, CA 94309, USA}
\author{E.K.~Pease} \affiliation{University of California Berkeley, Department of Physics, Berkeley, CA 94720, USA} \affiliation{Lawrence Berkeley National Laboratory, 1 Cyclotron Rd., Berkeley, CA 94720, USA} \affiliation{Yale University, Department of Physics, 217 Prospect St., New Haven, CT 06511, USA} 
\author{L.~Reichhart} \affiliation{Department of Physics and Astronomy, University College London, Gower Street, London WC1E 6BT, United Kingdom}  
\author{C.~Rhyne} \affiliation{Brown University, Department of Physics, 182 Hope St., Providence, RI 02912, USA}  
\author{S.~Shaw} \affiliation{University of California Santa Barbara, Department of Physics, Santa Barbara, CA 93106, USA} \affiliation{Department of Physics and Astronomy, University College London, Gower Street, London WC1E 6BT, United Kingdom} 
\author{T.A.~Shutt} \affiliation{Case Western Reserve University, Department of Physics, 10900 Euclid Ave, Cleveland, OH 44106, USA}  \affiliation{Kavli Institute for Particle Astrophysics and Cosmology, Stanford University, 452 Lomita Mall, Stanford, CA 94309, USA}
\author{C.~Silva} \affiliation{LIP-Coimbra, Department of Physics, University of Coimbra, Rua Larga, 3004-516 Coimbra, Portugal}  
\author{M.~Solmaz} \affiliation{University of California Santa Barbara, Department of Physics, Santa Barbara, CA 93106, USA}  
\author{V.N.~Solovov} \affiliation{LIP-Coimbra, Department of Physics, University of Coimbra, Rua Larga, 3004-516 Coimbra, Portugal}  
\author{P.~Sorensen} \affiliation{Lawrence Berkeley National Laboratory, 1 Cyclotron Rd., Berkeley, CA 94720, USA}  
\author{S.~Stephenson} \affiliation{University of California Davis, Department of Physics, One Shields Ave., Davis, CA 95616, USA}  
\author{T.J.~Sumner} \affiliation{Imperial College London, High Energy Physics, Blackett Laboratory, London SW7 2BZ, United Kingdom}  
\author{M.~Szydagis} \affiliation{University at Albany, State University of New York, Department of Physics, 1400 Washington Ave., Albany, NY 12222, USA}  
\author{D.J.~Taylor} \affiliation{South Dakota Science and Technology Authority, Sanford Underground Research Facility, Lead, SD 57754, USA}  
\author{W.C.~Taylor} \affiliation{Brown University, Department of Physics, 182 Hope St., Providence, RI 02912, USA}  
\author{B.P.~Tennyson} \affiliation{Yale University, Department of Physics, 217 Prospect St., New Haven, CT 06511, USA}  
\author{P.A.~Terman} \affiliation{Texas A \& M University, Department of Physics, College Station, TX 77843, USA}  
\author{D.R.~Tiedt} \affiliation{South Dakota School of Mines and Technology, 501 East St Joseph St., Rapid City, SD 57701, USA}  
\author{W.H.~To} \email{wto@slac.stanford.edu} \affiliation{Case Western Reserve University, Department of Physics, 10900 Euclid Ave, Cleveland, OH 44106, USA} \affiliation{SLAC National Accelerator Laboratory, 2575 Sand Hill Road, Menlo Park, CA 94205, USA}  \affiliation{Kavli Institute for Particle Astrophysics and Cosmology, Stanford University, 452 Lomita Mall, Stanford, CA 94309, USA} \affiliation{California State University Stanislaus, Department of Physics, 1 University Circle, Turlock, CA 95382, USA}
\author{M.~Tripathi} \affiliation{University of California Davis, Department of Physics, One Shields Ave., Davis, CA 95616, USA}  
\author{L.~Tvrznikova} \affiliation{University of California Berkeley, Department of Physics, Berkeley, CA 94720, USA} \affiliation{Lawrence Berkeley National Laboratory, 1 Cyclotron Rd., Berkeley, CA 94720, USA} \affiliation{Yale University, Department of Physics, 217 Prospect St., New Haven, CT 06511, USA} 
\author{S.~Uvarov} \affiliation{University of California Davis, Department of Physics, One Shields Ave., Davis, CA 95616, USA}  
\author{V.~Velan} \affiliation{University of California Berkeley, Department of Physics, Berkeley, CA 94720, USA}  
\author{J.R.~Verbus} \affiliation{Brown University, Department of Physics, 182 Hope St., Providence, RI 02912, USA}  
\author{R.C.~Webb} \affiliation{Texas A \& M University, Department of Physics, College Station, TX 77843, USA}  
\author{J.T.~White} \affiliation{Texas A \& M University, Department of Physics, College Station, TX 77843, USA}  
\author{T.J.~Whitis} \affiliation{Case Western Reserve University, Department of Physics, 10900 Euclid Ave, Cleveland, OH 44106, USA} \affiliation{SLAC National Accelerator Laboratory, 2575 Sand Hill Road, Menlo Park, CA 94205, USA} \affiliation{Kavli Institute for Particle Astrophysics and Cosmology, Stanford University, 452 Lomita Mall, Stanford, CA 94309, USA}
\author{M.S.~Witherell} \affiliation{Lawrence Berkeley National Laboratory, 1 Cyclotron Rd., Berkeley, CA 94720, USA}  
\author{F.L.H.~Wolfs} \affiliation{University of Rochester, Department of Physics and Astronomy, Rochester, NY 14627, USA}  
\author{J.~Xu} \affiliation{Lawrence Livermore National Laboratory, 7000 East Ave., Livermore, CA 94551, USA}  
\author{K.~Yazdani} \affiliation{Imperial College London, High Energy Physics, Blackett Laboratory, London SW7 2BZ, United Kingdom}  
\author{S.K.~Young} \affiliation{University at Albany, State University of New York, Department of Physics, 1400 Washington Ave., Albany, NY 12222, USA}  
\author{C.~Zhang} \affiliation{University of South Dakota, Department of Physics, 414E Clark St., Vermillion, SD 57069, USA}  

\collaboration{LUX Collaboration}\noaffiliation
\date{\today}
\begin{abstract}
\vspace*{2mm}
We present experimental constraints on the spin-dependent WIMP-nucleon elastic cross sections from the total 129.5 kg-year exposure acquired by the Large Underground Xenon experiment (LUX), operating at the Sanford Underground Research Facility in Lead, South Dakota (USA). A profile likelihood ratio analysis allows 90\% CL upper limits to be set on the WIMP-neutron (WIMP-proton) cross section of $\sigma_n$ = 1.6$\times 10^{-41}$~cm$^{2}$ ($\sigma_p$ = 5$\times 10^{-40}$~cm$^{2}$) at 35 GeV$c^{-2}$, almost a sixfold improvement over the previous LUX spin-dependent results. The spin-dependent WIMP-neutron limit is the most sensitive constraint to date.
\end{abstract}
\maketitle

\ifsection
\section{\label{sec:Introduction}Introduction}
\fi
The existence of dark matter is now supported by a wide array of astrophysical evidence, though the nature of its composition remains a mystery. The hypothetical WIMP (Weakly Interacting Massive Particle) is a compelling candidate, addressing both the observed astronomical phenomena as well as shortcomings of the Standard Model of particle physics (SM). The WIMP appears in many extensions of the SM, including supersymmetry~\cite{Jungman:1995df}, extra dimensions~\cite{Servant:2002aq}, and little Higgs theories~\cite{Birkedal:2006fz}. In these models, WIMPs may couple to SM particles mainly via scalar (spin-independent) and axial-vector (spin-dependent) interactions. The Large Underground Xenon (LUX) experiment, operating at the Sanford Underground Research Facility in Lead, South Dakota, is designed to detect such interactions through the scattering of galactic WIMPs with Xe nuclei. The LUX WIMP search program comprises two distinct exposures, termed WS2013 and WS2014--16. The combined dataset of both runs has been analyzed to produce world-leading limits on the spin-independent (SI) WIMP-nucleon cross section \cite{Akerib:2016vxi}. Here, we present the results for the spin-dependent (SD) coupling of WIMPs to protons and neutrons.

\ifsection
\section{\label{sec:LUXRun4}The LUX detector and WS2014--16 run}
\fi
LUX searches for WIMPs with a dual phase time projection chamber (TPC), detecting energy depositions through the resulting ionization and scintillation in the target material. The active detector volume, containing 250~kg of liquid xenon (LXe), is monitored by two horizontal arrays of 61 photomultiplier tubes (PMTs) each. The bottom array sits underneath a cathode grid in the LXe, while the top array looks on from above, in the gas phase. An energy deposition in the active region generates prompt scintillation photons as well as ionization electrons, which drift upwards under the influence of an applied electric field. The scintillation light is the first signal observed in the PMTs (S1). The second signal (S2) corresponds to the liberated charge: ionized electrons travel vertically to the liquid surface, where they are extracted into the gas phase and accelerated by a strong electric field. This produces additional vacuum ultraviolet (VUV) photons via electroluminescence. The S2 signal, originating close to the top PMT array, localizes the interaction in the $(x,y)$ plane. Additionally, the time delay between S1 and S2 gives the depth below the liquid surface, thereby allowing for full 3D position reconstruction. Position information is crucial for defining a fiducial volume, excluding background events that occur near the TPC walls. Further discrimination between WIMP signals (nuclear recoils, or NRs) and Compton or beta backgrounds (electron recoils, or ERs) is achieved using the S2 to S1 ratio.

As discussed in \cite{Akerib:2016vxi}, the recent WS2014--16 dataset was collected under substantially different detector conditions than WS2013: the electric drift field in the active volume featured spatial non-uniformities that evolved slowly over the course of the exposure. In particular, a significant radial component of the field was observed, as well as a vertical gradient in field magnitude. As a consequence of this field symmetry deformation, electron drift trajectories were bent radially inward, complicating the position reconstruction process. Though a similar phenomenon was seen in WS2013, the effects in WS2014--16 were more severe in magnitude, azimuthal distortion, and time-dependence. For example, in WS2014--16, an ionized electron originating near the edge of the cathode at a radius of $\sim$24~cm would reach the liquid surface at a radius of $\sim$10~cm (as opposed to $\sim$20~cm in WS2013). In addition to affecting electron drift paths, the field asymmetry introduced spatially-varying charge and light yields in the LXe. This is a result of the recombination physics of electron-ion pairs---more electrons (and thus fewer photons) will escape an interaction taking place in a region of greater field strength. As such, the boundaries of the bands populated by ERs and NRs in S1--S2 space vary slightly as a function of event position (and, to a lesser extent, calendar date).

A rigorous calibration regimen was established to address the challenges presented by the unique field geometry in the WS2014--16 analysis. Weekly $^{83\mathrm{m}}$Kr injections~\cite{Kastens:2009,Kastens:JINST,Manalaysay:2009yq}, in conjunction with periodic injections of tritiated methane\,\cite{Akerib:2015:tritium}, enabled the separation of electric field effects from the usual geometric effects typical of TPC detectors (i.e. spatial light collection efficiency and electron lifetime). Furthermore, $^{83\mathrm{m}}$Kr data were used to tune a 3-D electrostatic model of the detector, built with the \textsc{comsol Multiphysics} package \cite{comsolRef}. The electric field maps produced from this effort allowed for the time-dependent translation between true event position and the position inferred from  the observed S2.

NR calibrations were performed with neutrons from a deuterium-deuterium (DD) fusion generator \cite{Akerib:2015:dd, Verbus:2016sgw}. This technique, pioneered by LUX following the WS2013 run, was employed throughout the WS2014--16 exposure to monitor the detector's expected response to signal events. ER calibrations were obtained with tritiated methane, where the beta decays of tritium (endpoint 18.6~keV) give an excellent high statistics characterization of ER background  events~\cite{Akerib:2015:tritium}.

\begin{figure}
 \includegraphics[width=\columnwidth]{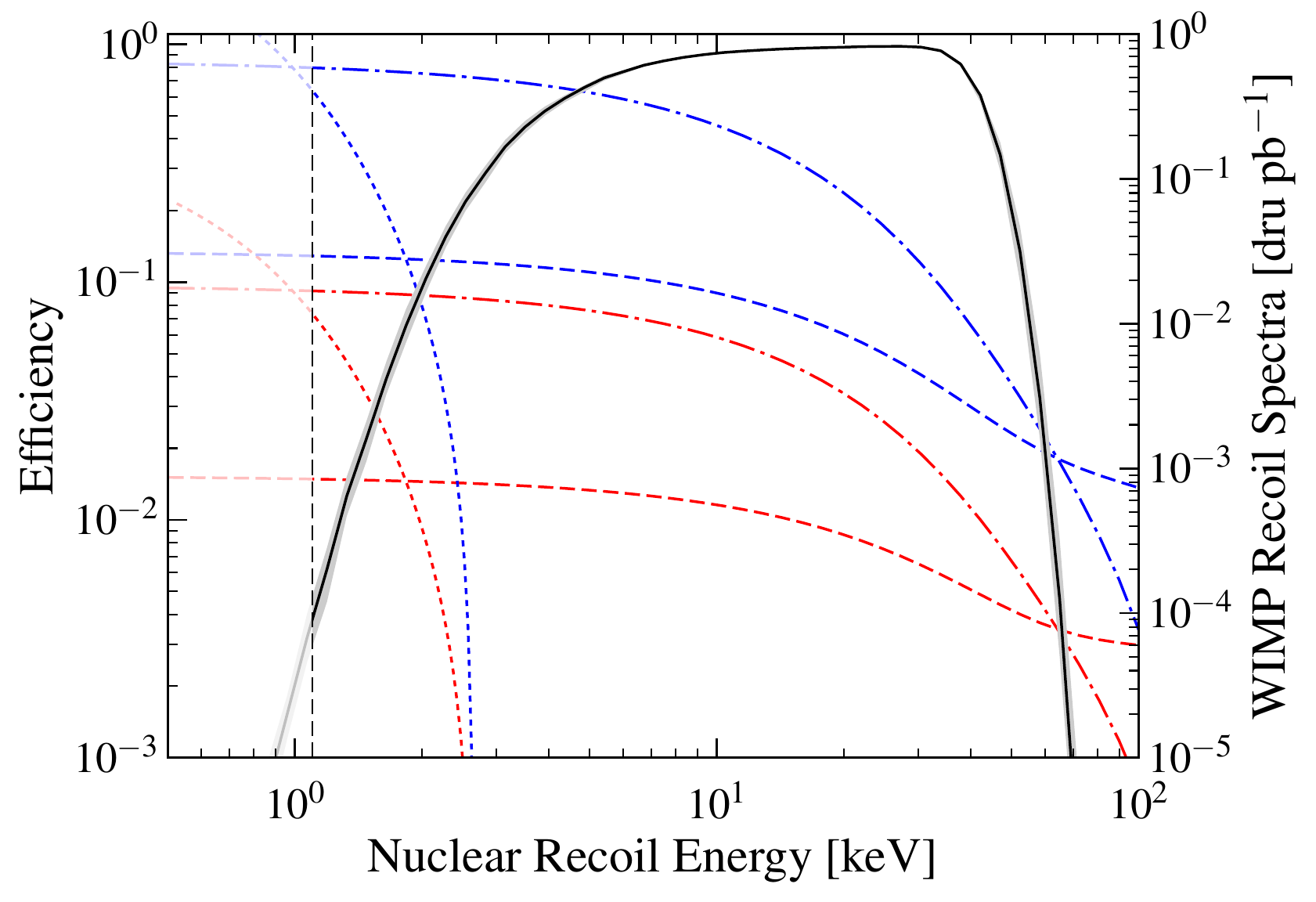}
 \caption{LUX total efficiency (black), averaged over the entire exposure. NR model uncertainties are illustrated by the gray band, indicating $\pm1\sigma$ variation. The vertical dashed line represents the analysis threshold of 1.1 keV (the lowest DD calibration point), below which we conservatively take the LUX efficiency to be zero. Sample WIMP spectra are plotted in color, with values corresponding to the right-hand y-axis. dru is the differential rate unit (events \SI{}{
 \kilo\gram^{-1} \day^{-1}\kilo eV^{-1}}), and the spectra have each been calculated with WIMP-nucleon cross section = 1 picobarn. Spectra for WIMP-proton and -neutron scattering are given by the red and blue curves, respectively, for three WIMP masses: 5 \GeVmass\,(dotted), 50 \GeVmass\,(dash-dotted), and 1000 \GeVmass\,(dashed). Structure functions from \cite{Klos} are used for this calculation.}
 \label{effPlot}
\end{figure}

\ifsection
\section{\label{sec:Dataset}The Dataset and Selection}
\fi
This analysis combines the WS2013 and WS2014--16 datasets in search of spin-dependent scattering between WIMPs and Xe nuclei. The WS2013 exposure was taken between April and August of 2013, totaling 95 live-days with a fiducial mass of 145 kg~\cite{Akerib:2015:run3}. A simple set of selection cuts were applied to the data, leaving 591 events in the region of interest. This dataset was previously analyzed to set SD WIMP-nucleon cross section limits~\cite{Akerib:2016:SD}. The WS2014--16 dataset was subjected to similar cuts, and furthermore featured a blinding protocol wherein fake WIMP events (``salt'') were injected into the data-stream. A full discussion of these data quality and selection cuts as well as the salting scheme can be found in Ref.\,\cite{Akerib:2016vxi}. In both runs, cuts were designed to select low-energy events with a single S1 followed by a single S2. The net effect on NR detection efficiency is illustrated in Fig.\ref{effPlot}, which shows the exposure-weighted efficiency of both WS2013 and WS2014--16 (black line, left axis scale). Efficiencies are calculated by applying analysis cuts to simulated NR events. Also plotted on the same energy scale are sample recoil spectra from SD WIMP-nucleon elastic scattering (right axis scale).

\ifsection
\section{\label{sec:SpinDepCoupling} Spin-dependent coupling between WIMPs and nucleons}
\fi
The spin-dependent coupling between WIMPs and sea quarks within target nucleons is evaluated using effective field theory (EFT) due to the non-perturbative nature of the strong force. As in \cite{Akerib:2016:SD}, we use the calculations made with 1-body (1b) and 2-body (2b) WIMP-quark scattering presented in \cite{Klos}. The differential cross section derived is
\begin{equation}
\frac{d\sigma}{dq^2} = \frac{8G_F^2}{(2J+1)v^2}S_A(q)\quad,
\end{equation}
where $G_F$ is the Fermi constant, $J$ is the ground state angular momentum of the nucleus, $v$ is the WIMP velocity, and $S_A(q)$ is the non-trivial momentum dependent axial-vector structure factor. In the zero momentum transfer limit it reduces to
\begin{multline}
S_A(0) = \frac{(2J+1)(J+1)}{4\pi J}\\
	\times|(a_p+\delta a_p)\langle \boldsymbol{\mathrm{S}}_p\rangle+(a_n+\delta a_n)\langle\ \boldsymbol{\mathrm{S}}_n\rangle|^2\quad.
\end{multline}
The $a_{p,n}$ are WIMP-proton(neutron) coupling constants for 1b currents and $\delta a_{p,n}$  account for the effects of 2b currents (in the nomenclature of \cite{Klos}, $\delta a_{p,n} = \pm a_1 \delta a_1(0)$). $\langle \boldsymbol{\mathrm{S}}_{p,n}\rangle$ are the spin expectation values of the proton and neutron groups in the nucleus. The case of ``proton-only'' coupling ($a_n = 0$) is so named because, in the 1b regime, only the protons contribute to $S_A(0)$ (the same applies to ``neutron-only'' with $a_p = 0$). However, with the introduction of 2b currents, neutrons may be involved in a ``WIMP-proton'' scattering event, changing this picture. Thus, for target nuclei with unpaired neutrons such as $^{129}$Xe and $^{131}$Xe, sensitivity is much greater in the neutron-only case (since $\langle \boldsymbol{\mathrm{S}}_{n}\rangle \gg \langle \boldsymbol{\mathrm{S}}_{p}\rangle$), though 2b currents allow for non-zero sensitivity to a proton coupling. $^{129}$Xe and $^{131}$Xe occur naturally in xenon with respective abundances of 29.5\% and 23.7\%.

The non-zero momentum transfer structure factors $S_A(q)$ can be decomposed as
\begin{equation}
\begin{split}
S_p(q) & = S_{00}(q)+S_{01}(q)+S_{11}(q)\\
S_n(q) & = S_{00}(q)-S_{01}(q)+S_{11}(q)\quad,
\end{split}
\end{equation}
where $S_{ij}(q)$ are the isovector/isoscalar components. Fit parameters for these functions are listed for $^{129}$Xe and $^{131}$Xe in Table IV of \cite{Klos}. Using these, we can compute $d\sigma/dE$ and generate various SD WIMP-nucleon recoil spectra (see Fig.\,\ref{effPlot}). We assume a standard Maxwellian WIMP velocity distribution near Earth with $v_0=220$\,km/s, $v_{esc}$ = 544\,km/s, and $\rho_0$ = 0.3\,GeV/cm$^3$. To calculate the average relative Earth velocity during the exposure, we follow Ref.\cite{Savage:2006qr} to obtain $\langle v_{obs} \rangle_{\mathrm{WS2013}}=245$\,km/s and $\langle v_{obs} \rangle_{\mathrm{WS2014-16}}= 230$\,km/s.

\ifsection
\section{\label{Modelling}Statistical Analysis}
\fi
A two-sided profile likelihood ratio (PLR) statistic is used to test signal hypotheses \cite{Cowan:2010js}, whereby the complete LUX dataset is compared against a multi-channel, extended, unbinned likelihood function \cite{Cranmer:2015nia,CMS-NOTE-2011-005}. LUX data are categorized into five ``channels'': one corresponds to the WS2013 exposure, and the remaining four represent discrete time periods of relatively constant detector conditions in the WS2014--2016 dataset, termed ``date bins.'' The simultaneous model is thus the product of each channel's likelihood, along with the nuisance parameter constraints. Nuisance parameters, representing systematic uncertainties in the model, are described in \cite{Akerib:2016vxi}, as are the components of the background model. Here, we review some details of the model construction.

The signal and background probability distribution functions (PDFs) for WS2013 are defined in four observables: corrected interaction radius and height, $\emph{S1}$, and $\log_{10}(\emph{S2})$\,\cite{Akerib:2015:run3}. Uncorrected S2 position coordinates $\mathrm{\{ r_{S2} ,\, }\phi\mathrm{_{S2} , z_{S2} \} }$ are used in WS2014--16, with the loss of axial symmetry necessitating the introduction of the third spatial dimension. ER backgrounds as well as the NR signal are modeled by further subdividing the data into segments of drift time.\footnote{A small NR background from $^8$B solar neutrinos is also modeled with this technique. Neutrons (muon-induced or from detector components) can also produce NR background events, though the estimated rate is negligible\,\cite{Akerib:2016vxi,Akerib:2014:bg}.} For each date bin of WS2014--16, calibration data is used to tune $\mathrm{\{}\emph{S1}, \emph{S2}\}$ response models in four horizontal slices of the detector (within which the field strength variation is acceptably low). From these date- and depth-specific models, implemented with the Noble Element Simulation Technique (NEST)\,\cite{Lenardo:2014}, Monte Carlo (MC) data are generated to produce 16 ER and NR PDFs. Spatial PDFs are built separately using MC from the \textsc{Geant4}-based\,\cite{Agostinelli:2002hh} \textsc{LUXSim}\,\cite{Akerib:2012:luxsim} software: simulated event positions are transformed into the observed S2 coordinate space via the $^{83\mathrm{m}}$Kr-derived field maps, once for each date bin.

As in WS2013, the WS2014--16 data contains a background population that defies the ER and NR description. Interactions occurring very near the TPC walls suffer charge loss to the PTFE panels, suppressing the S2 signal. Since position reconstruction statistical uncertainty scales as S2$^{-1/2}$, these low charge yield events are more likely to be mis-reconstructed as taking place within the fiducial volume (because of the long tail in their radial distribution). An empirical model is constructed to describe this population using control samples of the dataset outside the region of interest. More so than the other models, this ``wall'' model features strong correlations between position and pulse area observables. For example, the width of the radial distribution is dependent on uncorrected S2, which is itself a function of the  corrected $\emph{S2}$ and $\mathrm{z_{S2}}$ observables used in the PLR. Furthermore, in observed S2 position coordinates, the radial position of the wall varies with $\{\phi\mathrm{_{S2} , z_{S2} }\}$. The final PDF is implemented as a finely binned 5-dimensional histogram in each date bin, via an extension of the technique described in Ref.\,\cite{Lee:2015}. Specifics of the model construction will be detailed in a forthcoming publication.

The full background model is found to be a good fit to the combined dataset. The data are consistent with the background-only hypothesis (PLR $p = 0.35$) when testing a \SI{50}\,\GeVmass\,signal. As a further cross-check, the WS2013 and WS2014--16 PDFs are separately projected into 1-dimensional spectra for each observable. These are compared to data with a Kolmogorov-Smirnov test, demonstrating acceptable goodness of fit ($p \geq .05$ and $p \geq .6$ in WS2013 and WS2014--16, respectively)\, \cite{Akerib:2015:run3,Akerib:2016vxi}. Finding no evidence for WIMP signals in the data, we proceed in setting 90\% confidence level (CL) limits on the WIMP-nucleon cross section, in the case of spin-dependent coupling.

\ifsection
\section{\label{Results}{Results}}
\fi
For a given WIMP mass and choice of coupling type, the PLR test statistic distribution is constructed at a range of signal cross sections from MC pseudo-experiments generated with the R\textsc{oo}S\textsc{tats} package \cite{2010acat.confE..57M}. The $p$-value  of the observed data is then calculated over this range, where by definition the 90\% CL upper limit is given by the cross section at which $p=0.1$. In using the raw PLR test statistic, however, an experiment may benefit unreasonably from background under-fluctuations. To safeguard against setting an upper limit at a cross section to which LUX is insensitive, a power constraint \cite{Cowan:2011an} is imposed at $-1\sigma$ of the expected sensitivity calculated from background-only trials (as in \cite{Akerib:2016vxi}). Since the WS2013 limits were reported with an overly conservative power constraint at the median expected sensitivity, this combined result exhibits a stronger improvement than suggested simply by the increase in exposure.

\begin{figure}
 \includegraphics[width=\columnwidth]{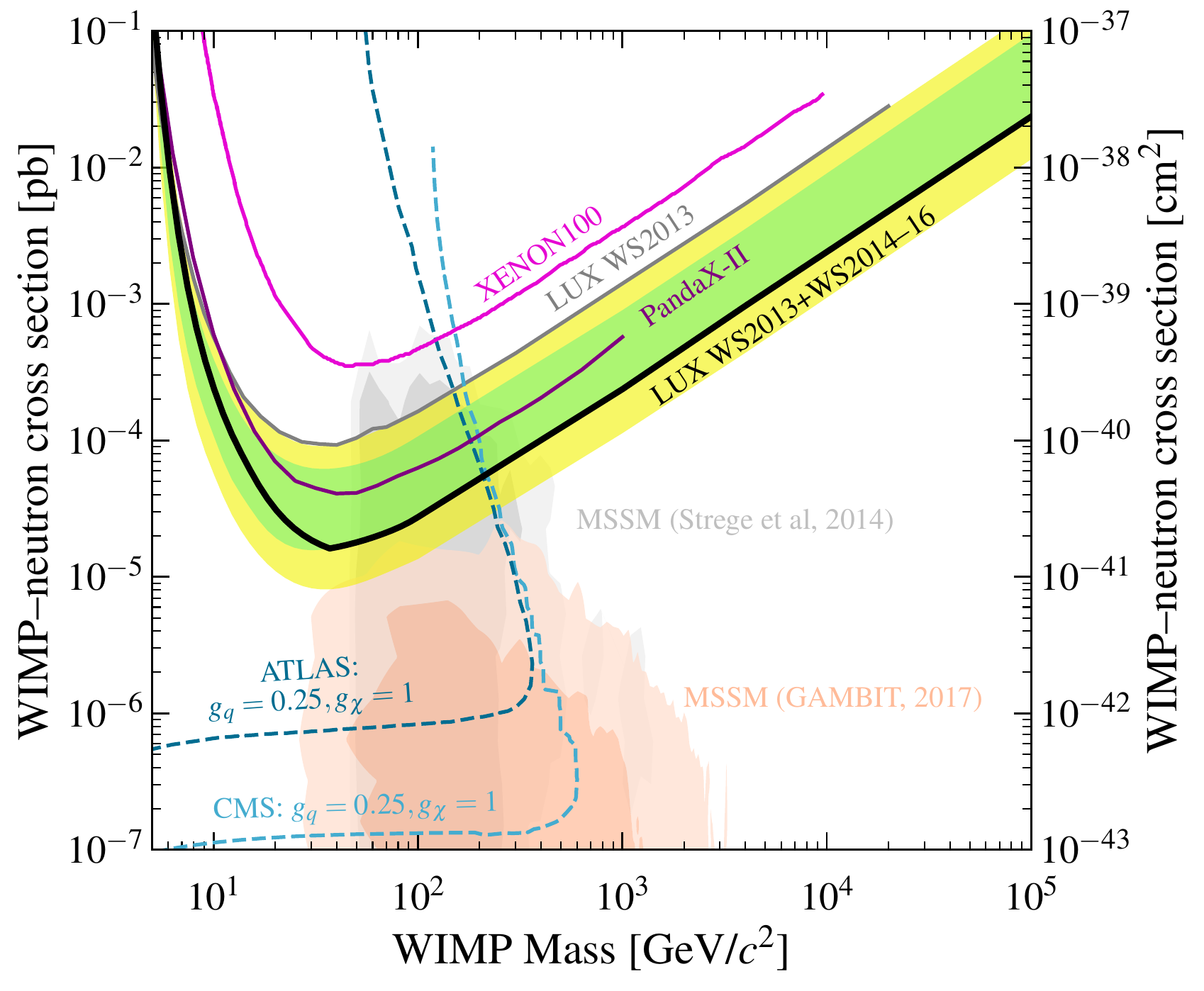}
 \includegraphics[width=\columnwidth]{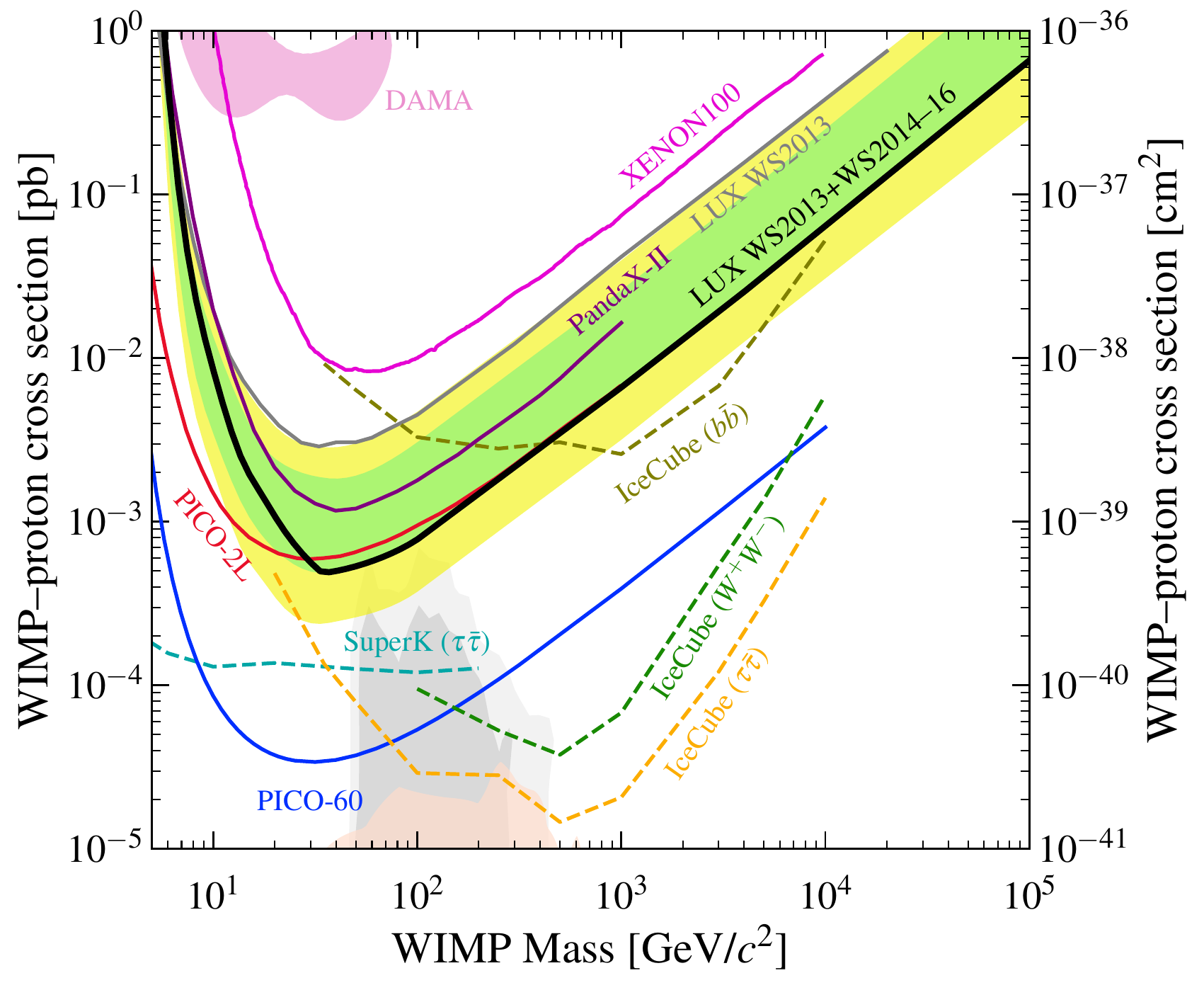}
 \caption{90\% CL upper limits on WIMP-neutron (top) and WIMP-proton (bottom) cross section. Results from this analysis are shown in thick black (``LUX WS2013+WS2014--16''), with the range of expected sensitivity indicated by the green (1-$\sigma$) and yellow (2-$\sigma$) bands. Solid gray curves show the previously published LUX WS2013 limits\,\cite{Akerib:2016:SD}. Constraints from other LXe TPC experiments are also shown, including XENON100\,\cite{Aprile:2013doa} and PandaX-II\,\cite{Fu:2016ega}. In the top panel, model-dependent (axial-vector mediator with indicated couplings) LHC search results are represented by dashed lines, with CMS\,\cite{Sirunyan:2017hci} in light blue, and ATLAS\,\cite{Aaboud:2017dor} in dark blue. As calculated by a new profile likelihood scan of the MSSM7\,\cite{Athron:2017yua}, favored parameter space is shown as dark (1-$\sigma$) and light (2-$\sigma$) peach regions; an earlier calculation using the MSSM-15\,\cite{Strege:2014ija} is shown in gray, with analogous shading of confidence levels.
In the bottom panel, the DAMA allowed region (as interpreted in \cite{SavageDAMA}) is shown in pink (the analogous neutron-only region is above the bounds of the plot). Such an interpretation is in severe tension with this result, as well as the PICO-2L\,\cite{Amole:2016pye} and PICO-60\,\cite{Amole:2017dex} constraints. Selected limits from indirect searches at neutrino observatories (Super-Kamiokande\,\cite{Choi:2015ara} and IceCube\,\cite{Aartsen:2016zhm}) are plotted as dashed lines.}
\label{limits}
\end{figure}

\begin{figure}
\includegraphics[width=.83\columnwidth]{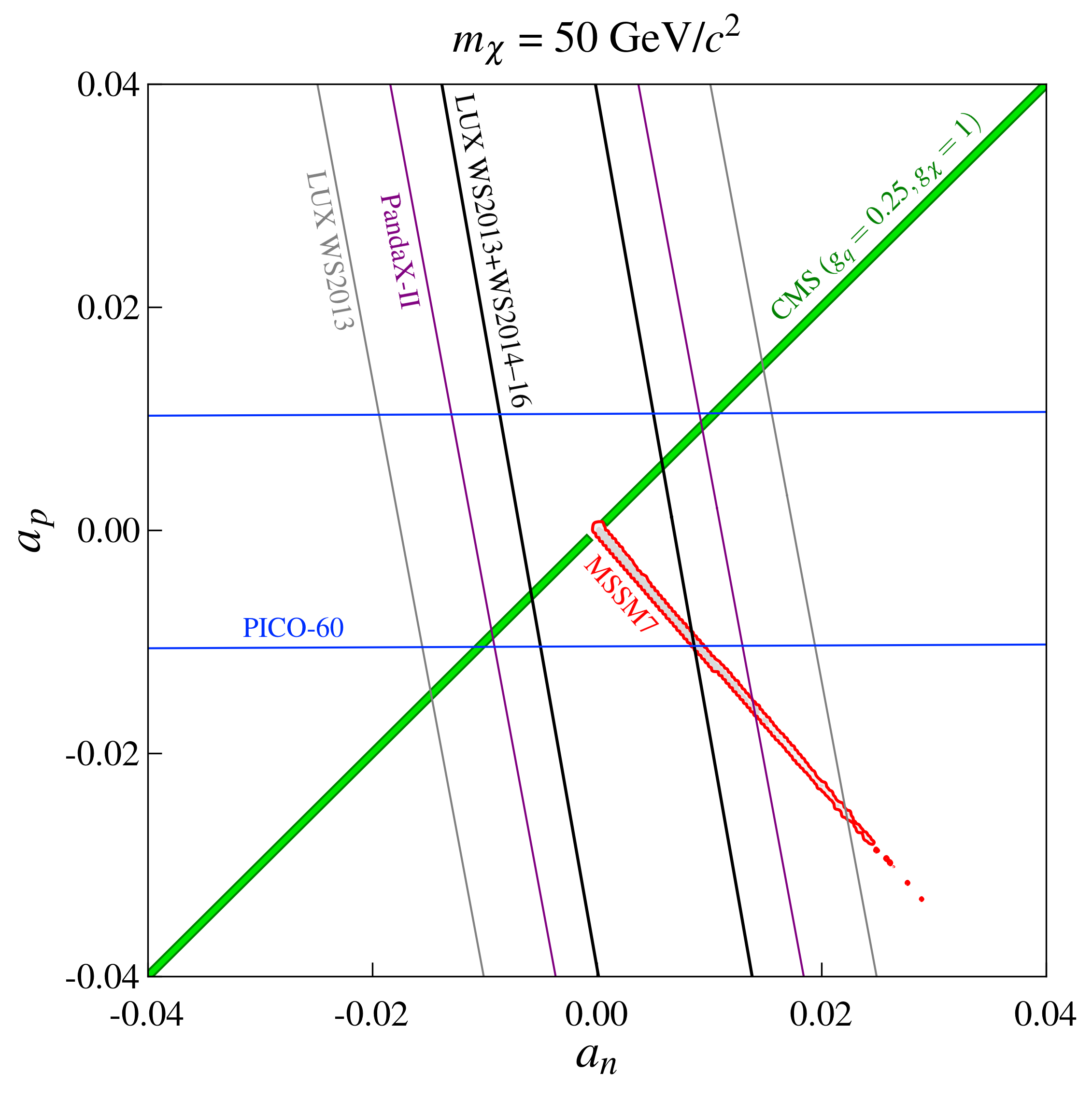} 
\includegraphics[width=.83\columnwidth]{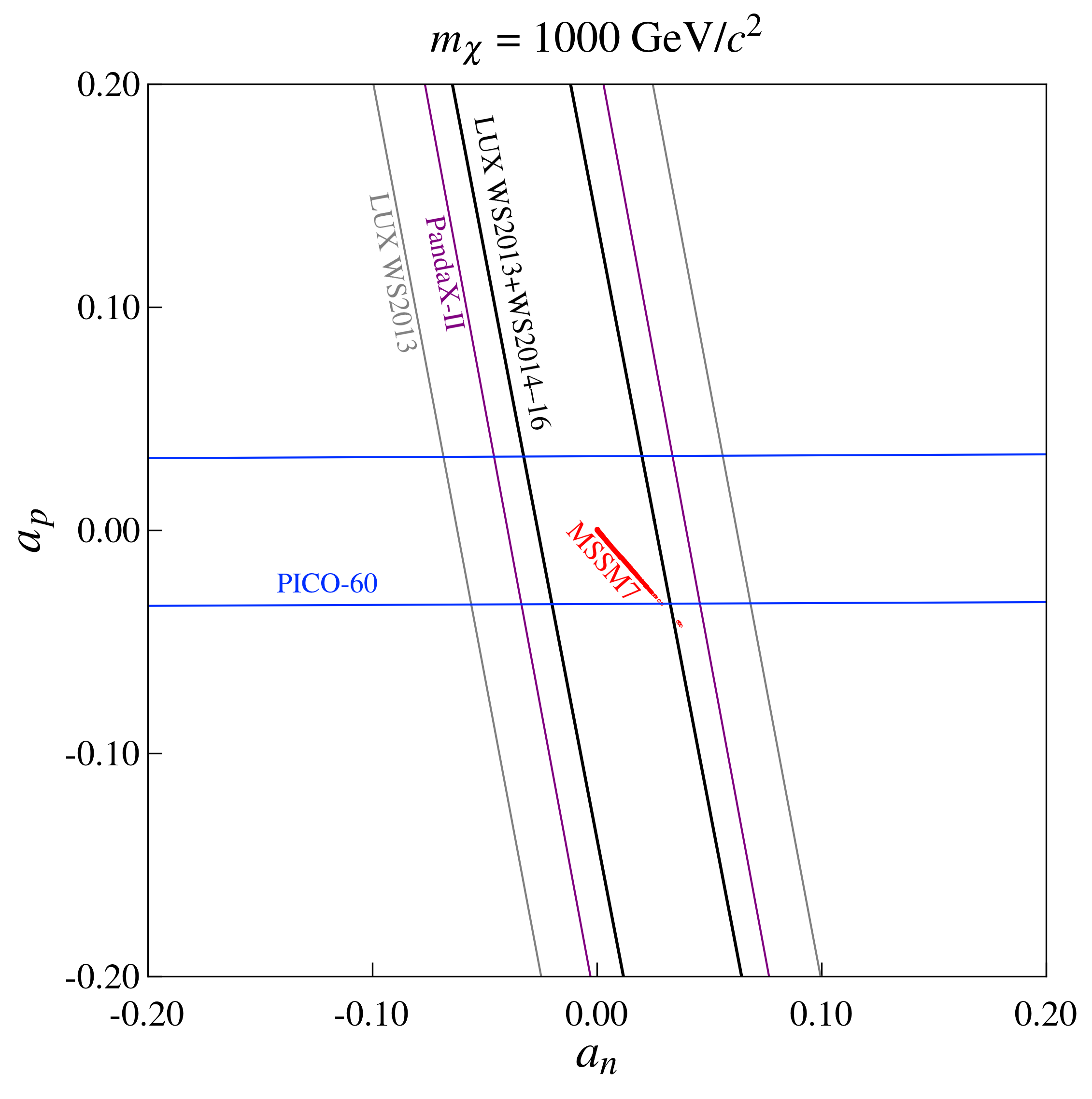}
\vspace*{-3mm}
 \caption{90\% CL exclusions on coupling parameters $a_n$ and $a_p$ for 50~\GeVmass\,and 1000~\GeVmass\,WIMPs. Ellipse boundaries are colored as in Fig.\,\ref{limits}\,: this result (thick black), LUX WS2013 (gray), PandaX-II (purple), and PICO-60 (blue). Geometrically, Eq.\,\ref{eq:tovey} describes a rotated ellipse when the sum is performed over multiple isotopes with distinct $\sigma_p^{A}/\sigma_n^{A}$, as is the case for LXe experiments. PICO-60 considers only $^{19}$F (for which $\langle \boldsymbol{\mathrm{S}}_{n}\rangle \! \sim \! 0$), and thus sets limits only on $a_p$. The innermost region (bounded by LUX and PICO-60) represents parameter space not in tension with experimental data. The model-dependency of the LHC results is apparent in this plane, as the CMS excluded region (shown as a green band) is restricted to the $a_n = a_p$ line (see main text for important caveat). This line is absent from the lower panel since, in this treatment, CMS is insensitive to WIMPs at the TeV mass scale. MSSM7 favored regions from the GAMBIT scan are also shown, with a red contour at the 2-$\sigma$ level for visibility. The degeneracies assumed in the MSSM7 Lagrangian lead to the tight correlation between $a_n$ and $a_p$. This scan includes a range of possible WIMP masses (unlike the mass-specific experimental exclusions), and thus appears identically in each panel, noting the change in axis scale. Additionally, the scans include models with sub-dominant relic densities, for which experimental limits are rescaled accordingly.}
\label{an_vs_ap}
\end{figure}

The advance in sensitivity can be seen in Fig. \ref{limits}\,, which shows cross section limits as a function of WIMP mass in the cases of neutron- and proton-only coupling. The limits from the combined LUX data are plotted as a thick black line, labeled ``LUX WS2013+WS2014--16''. LUX is more sensitive to the neutron-only scenario, owing to the unpaired neutron in $^{129}$Xe and $^{131}$Xe nuclei, and sets a minimum upper limit of \SI{1.6e-41}{\centi\metre\squared} at \SI{35}\,\GeVmass, a nearly sixfold improvement over the previous WS2013 result. Indeed, among direct detection experiments, LUX is world-leading in sensitivity to WIMP-neutron interactions. Also shown are sample results from LHC searches, interpreted as exclusions in the WIMP mass vs. cross section plane by assuming mediator coupling parameters in a $Z'$-like simplified model\cite{Buchmueller2015,Boveia:2016mrp}. Though strictly model-dependent, these limits present strong constraints below $\sim$500 GeV, whereas the sensitivity of LXe TPCs extends to much higher WIMP masses.

To contextualize these WIMP-neutron cross section limits, regions of favored parameter space derived from a 7-parameter Minimal Supersymmetric Standard Model (MSSM7\,\cite{Bergstrom:1995cz}) are also indicated. These regions, newly calculated \cite{Athron:2017yua} by the GAMBIT collaboration \,\cite{Athron:2017ard, Athron:2017qdc, Workgroup:2017lvb, Workgroup:2017bkh}, are generated from scans of the MSSM7, where constraints from a suite of experimental results appear in the likelihood functions. In particular, recent results from LUX\,\cite{Akerib:2016vxi} and PandaX--II\,\cite{Tan:2016zwf} are included. As such, the favored parameter space is appropriately just beyond the sensitivity of this work (since the dataset used here is the same as in the SI analysis of Ref.\,\cite{Akerib:2016vxi}, which is already taken into account by the GAMBIT profile likelihood scan). Another region of favored parameter space from a 2014 scan of MSSM-15\,\cite{Strege:2014ija} is shown for comparison, illustrating the rapid advance of the field and the contribution of direct detection searches such as LUX.

In the proton-only scenario, high mass limits from this result now coincide with those previously set by the PICO-2L experiment. The recent limit from PICO-60 sets the standard for proton-only sensitivity in direct detection, bolstering the constraints from indirect searches performed by the neutrino detectors IceCube and Super-Kamiokande. CMS and ATLAS take $g_q$ (i.e. the coupling of quark type $q \in \{u, d, s\ldots\}$ to the axial-vector mediator) to be universal, and thus set equivalent limits on WIMP-neutron and -proton cross section (the curves are omitted in the bottom panel of Fig.\,\ref{limits} for clarity). However, we note that in a more careful treatment of the simplified model, renormalization group evolution of the couplings from the LHC to nuclear energy scale leads to significant isospin violation (see Ref.\,\cite{Crivellin:2014qxa,DEramo:2014nmf,DEramo:2016gos}).

The cases of neutron- and proton-only coupling fall on the axes of the more general parameter space spanned by $a_n$ and $a_p$. By following the prescription laid out in \cite{Tovey:2000mm}, elliptical exclusions in this plane are made according to:
\begin{equation}
\label{eq:tovey}
\sum_A \left( \frac{a_p}{\sqrt{\smash[b]{\sigma_p^{A}}}} \pm \frac{a_n}{\sqrt{\smash[b]{\sigma_n^{A}}}} \right)^2 > \frac{\pi}{24 G_F^2 \mu_p^2}\quad,
\end{equation}
where the sum is performed over target isotopes with mass numbers $A$, and $\sigma_{p(n)}^{A}$ are the 90\% CL upper limits on WIMP-proton(neutron) cross-section, calculated individually from these isotopes. For the PICO-60 results, where only the proton-only results are reported, limits are calculated according to \cite{PhysRevD.71.123503}. Exclusions are shown in Fig.\,\ref{an_vs_ap} for two choices of WIMP mass, highlighting the complementary experimental reach of LXe and fluorine-rich detectors. The CMS results are also shown in this plane as exclusions along the $a_n = a_p$ line (since $g_q$ is assumed to be the same for all quarks)\,\cite{Sirunyan:2017hci,McCabe2017}. Results from the GAMBIT scans of the MSSM7 are also displayed.

\ifsection
\section{\label{Conclusion}Conclusion}
\fi
In conclusion, the complete LUX dataset has been analyzed to set limits on SD WIMP-nucleon scattering. World-leading constraints are presented for neutron-only coupling, complementing searches for particle production at the LHC. Further complementarity with the PICO-60 result is achieved in the 2D $a_n$--$a_p$ plane. Future work will investigate a more complete set of EFT interaction operators, beyond those that define the standard SI and SD paradigm.

\ifsection
\begin{acknowledgments}
\fi
The authors thank C. McCabe for useful discussions on the interpretation of LHC searches, and P. Scott and the GAMBIT collaboration for providing their results on the $a_n$--$a_p$ plane. We also thank F. D'Eramo, B. J. Kavanagh, and P. Panci for pointing out subtleties that arise from the running of couplings in simplified dark matter models. This work was partially supported by the U.S. Department of Energy (DOE) under award numbers DE-AC02-05CH11231, DE-AC05-06OR23100, DE-AC52-07NA27344, DE-FG01-91ER40618, DE-FG02-08ER41549, DE-FG02-11ER41738, DE-FG02-91ER40674, DE-FG02-91ER40688, DE-FG02-95ER40917, DE-NA0000979, DE-SC0006605, DE-SC0010010, and DE-SC0015535; the U.S. National Science Foundation under award numbers PHY-0750671, PHY-0801536, PHY-1003660, PHY-1004661, PHY-1102470, PHY-1312561, PHY-1347449, PHY-1505868, and PHY-1636738; the Research Corporation grant RA0350; the Center for Ultra-low Background Experiments in the Dakotas (CUBED); and the South Dakota School of Mines and Technology (SDSMT). LIP-Coimbra acknowledges funding from Funda\c{c}\~{a}o para a Ci\^{e}ncia e a Tecnologia (FCT) through the project-grant PTDC/FIS-NUC/1525/2014. Imperial College and Brown University thank the UK Royal Society for travel funds under the International Exchange Scheme (IE120804). The UK groups acknowledge institutional support from Imperial College London, University College London and Edinburgh University, and from the Science \& Technology Facilities Council for PhD studentships ST/K502042/1 (AB), ST/K502406/1 (SS) and ST/M503538/1 (KY). The University of Edinburgh is a charitable body, registered in Scotland, with registration number SC005336.

This research was conducted using computational resources and services at the Center for Computation and Visualization, Brown University, and also the Yale Science Research Software Core. The $^{83}$Rb used in this research to produce $^{83\mathrm{m}}$Kr was supplied by the United States Department of Energy Office of Science by the Isotope Program in the Office of Nuclear Physics.

We gratefully acknowledge the logistical and technical support and the access to laboratory infrastructure provided to us by SURF and its personnel at Lead, South Dakota. SURF was developed by the South Dakota Science and Technology Authority, with an important philanthropic donation from T. Denny Sanford, and is operated by Lawrence Berkeley National Laboratory for the Department of Energy, Office of High Energy Physics.

\ifsection
\end{acknowledgments}
\fi

\bibliography{main}

\end{document}